\documentstyle[11pt]{article}
\newcommand{\abz}{\vskip10pt\noindent}
\def \rp { \ \rho_{\varphi} \ }
\def \pp { \ p_{\varphi} \ }

\def \be  {\begin{equation}}
\def \ee  {\end{equation}}

\begin{document}
\begin{titlepage}
\[ \]
\begin{center}
{\Large  Inflationary cosmology with two-component
             fluid and thermodynamics}

\[ \]
\[ \]
\[ \]
Edgard Gunzig
\footnote{Universit\'{e} Libre de Bruxelles, Service de
Chimie-Physique,
Campus Plaine CP~231, 1050 Bruxelles, Belgium}, \
Alexei V. Nesteruk  \ and
Martin Stokley
\footnote{School of Computer Science and
Mathematics, Portsmouth University, PO1 2EG England}
\end{center}
\[ \]
\[ \]
{\em Running head:}   Inflationary cosmology with two-component
             fluid and thermodynamics
\[ \]
\[ \]
\[ \]
{\em Correspondence:}
\[ \]
{\sf A V Nesteruk\\
School of Computer Science and Mathematics\\
Portsmouth University\\
Mercantile House\\
Hampshire Terrace\\
Portsmouth  PO1 2EG\\
England
\[ \]
TEL:   + (01705) 843 108\\
FAX:   + (01705) 843 106\\
EMAIL: alexei.nesteruk@port.ac.uk}
\end{titlepage}
\[ \]
\[ \]
{\bf Abstract}
\[ \]
\[ \]
We present a simple and self-consistent cosmology with
a phenomenological model of quantum creation of radiation and
matter due to  decay of the cosmological constant $\Lambda$.
The decay drives a non-isentropic inflationary epoch, which
exits smoothly to the radiation-dominated era, without reheating,
and then evolves to the dust era. The initial vacuum for radiation  and
matter is a regular Minkowski vacuum. The created radiation and
matter obeys standard thermodynamic laws, and the total entropy
produced is consistent with the accepted value. This paper is an
extension of the model with the decaying cosmological constant
considered in \cite{gmn}. We  compare our model with the quantum
field theory approach  to creation of particles in curved space.
\[ \]

\section{Introduction}

The aim of this paper is to extend the scenario of the evolution of
the universe with smooth exit from inflation, and particle production at

the expense of the decaying cosmological constant, developed in
paper \cite{gmn}. In that paper thermodynamics and Einstein's
equations led to an equation in which the Hubble rate $H$ is determined
by the particle number $N$. The model is completed by specifying
the particle creation rate $\Gamma=\dot{N}/N$, which led to a
second-order
evolution equation for $H$.
The evolution equation for $H$ then has a simple exact
solution, in which a non-adiabatic inflationary era exits smoothly
to the radiation era, without a reheating transition.
For this solution, there were given  exact expressions
for the cosmic scale factor, energy density of radiation and vacuum,
temperature and entropy.

In the paper \cite{infl98} we  generalised the abovementioned
results for the case of a scalar field $\varphi$ interacting with
radiation via the gravitational field, leading to
cosmological evolution with  smooth exit from inflation.
Our particular task was to determine whether the theory formulated in
terms of the scalar field $\varphi$ could lead to any new results in
comparison with the previous case of a decaying cosmological
constant. We  concluded that the presence of the scalar field
$\varphi$ in this model did not change considerably the physical
results which have been obtained in the paper  \cite{gmn}.
As result we argued that  models with  decaying cosmological
constant $\Lambda$ corresponding to a special case of the equation
of state $\pp = - \rp$,   describe adequately the smooth transition
from inflation to radiation and give a reasonable prediction
for the entropy of matter in the universe.

In this paper we continue to argue along the lines  of
the paper  \cite{gmn} generalising its results for a two-component
cosmological fluid, i.e. radiation and matter. The aim of the paper
is to build a scenario of overall evolution of the universe
with smooth exit from inflation to the radiation dominated stage and
then
its further evolution to  the matter-dominated universe.

The background of this paper is constituted by two ideas in
physical cosmology. One of them is connected with the longstanding
attempt to explain all matter in the universe as produced by
quantum creation from vacuum. This has been studied via quantum
field theory in curved spacetime
(see for example \cite{hupark77,birdav,hu,creation,nestott}).
Most cosmological models exhibit a singularity which
presents  difficulties  for interpreting quantum effects, because
all  macroscopic parameters of created particles are infinite
there.  This leads to the  problem of the initial vacuum
(see discussion in \cite{gmn}). One attempt to
overcome these problems is via
incorporating the effect of particle creation
into Einstein's field equations. For example, in
the papers of the Brussels group \cite{edgard1},
the quantum effect of particle creation  is considered
in the context  of the thermodynamics of open systems,
where it is interpreted as an additional negative
pressure, which emerges from a re-interpretation of
the energy-momentum tensor.
This effect is irreversible in the sense that spacetime
can produce matter, leading to growth of entropy,
while the reverse process is thermodynamically
forbidden. These results were generalized in  a covariant
form in \cite{lima1}. Our approach differs from that of
\cite{edgard1,lima1} in that we do not modify the field
equations. Instead, we associate  the source of created particles
as a decaying cosmological constant $\Lambda$.

A number of decaying vacuum models has appeared in the literature
(see \cite{lima2,vl} and references cited
there). A review of the different phenomenological
models  of  evolution with variable cosmological term
can be  found in  paper \cite{overcoop}.

Inflationary models with fixed cosmological constant and cold dark
matter have been successful in accounting for the microwave
background and large-scale structure observations, while also
solving the age problem (see \cite{krausturner}). However, these
models are challenged by the reduced upper limits on $\Lambda$
arising from the Supernova Cosmology Project (see \cite{cc}),
and also by the long-standing problem of reconciling the very large
early-universe vacuum energy density with the very low
late-universe limits \cite{vl}.

One resolution of these problems  is a decaying cosmological
constant $\Lambda$ which is treated as a  dynamical parameter.
This approach was typical for the quantum field theorists for
many years (see for example \cite{qft}).
Many potential sources of fluctuating
vacuum energy have been identified which were to give rise to
a negative energy
density which grows with time, tending to cancel out any pre-existing
positive cosmological term and drive the net value of $\Lambda$ toward
zero. Processes of this kind are among the most promising ways to
resolve the longstanding cosmological `constant' problem
(see \cite{wei89} for review).
It is worth mentioning the recent paper of Parker \cite{parker}
indicating an attempt to revive  the idea of
the cosmological constant as a purely quantum effect associated with
the renormalization of the general relativistic action.

In ad hoc
prescriptions, the functional form of $\Lambda(t)$ or
$\Lambda(a)$ or $\Lambda(H)$ (where $a$ is the scale factor and
$H$ is the hubble rate) is effectively assumed a-priori
(see the review \cite{overcoop} where all known forms
for $\Lambda$ are listed).
Typically, the solutions arising from ad hoc prescriptions for
$\Lambda$ are rather complicated, and moreover, it is often
difficult to provide a consistent simple interpretation of the
features of particle creation, entropy and thermodynamics.

In contrast to many other models, we propose a simple, exact and
thermodynamically consistent cosmological history. The latter
originates from a regular initial vacuum.
Together with  naturally defined asymptotic conditions for the
number of created particles this leads to a simple expansion law and
thermodynamic properties, and to a definite estimate for the
total entropy in the universe. Since the exit from inflation to
the radiation era is smooth, we avoid the problem of matching at
the transition. A similar smooth evolution has been used in
\cite{gmn,infl98,roydavid,alexei_96,caldwell}.

The choice of $a$ as dynamical variable and the very simple form of
$H(a)$ that meets the physical conditions, lead to elegant expressions
for all parameters describing the radiation and decaying vacuum, and
also to a physically transparent interpretation of these results,
including the estimate of entropy.

We use units with $8\pi G$, $c$ and $k_{\rm B}$ equal 1.

\section{Model}
\abz
{\it Metrics and Matter}
\abz
We consider a spatially flat FRW universe we has the metric
\begin{equation}
ds^2 = - dt^2 + a^2(t) \left[dx^2 + dy^2 + dz^2\right]\,,
\label{metric}
\end{equation}
containing a uniform two-component cosmological fluid, which is composed
of non-interacting
matter and radiation. The radiation has an energy
density $\rho_{r}(t)$ and a pressure $p_r = \rho_{r}/3$ while the
non-relativistic matter
has an energy density $\rho_{m}(t)$ and pressure $p_m=0$.
( we assume that $\rho_m \ll \rho_r, \, n_m \ll n_r$ at high temperature

$T$, so that  we can neglect $p_m=n_m T$ with respect to $p_r$).

The energy momentum tensor of these components correspondingly is
\[
T^R_{\mu\nu}={1\over3}\rho_r(t)\left[4u_\mu u_\nu+g_{\mu\nu}\right]\,,
\qquad
T^M_{\mu\nu}=\rho_m(t) u_\mu u_\nu\,,
\]
\abz
{\it Decaying vacuum}
\abz
We also consider matter corresponding to the quantum
vacuum energy,  with energy momentum tensor
\[
T^{Q}_{\mu\nu} \equiv \langle \widehat{T}^Q_{\mu\nu} \rangle =
\left[{\Lambda (t)\over 8 \pi G}\right]
g_{\mu\nu}\,.
\]
(We shall use units such that $8\pi
G= 1 = c$ henceforth)  The conservation equations
$\nabla^\nu(T^R_{\mu\nu} + T^M_{\mu\nu} + T^Q_{\mu\nu})=0$
reduce to
\begin{equation}
\dot{\rho_r} +  \dot{\rho_m}+  4 H \rho_r + 3 H \rho_m = -
\dot{\Lambda}\,,
\label{conserv}
\end{equation}
(which is a special form of the first law of thermodynamics),
where $H=\dot{a}/{a}$ is the Hubble rate. We can rewrite this equation,
introducing an enthalpy of radiation and  matter
\[
h = h_r + h_m; \quad h_r \equiv \rho_r + p_r = 4/3 \, \rho_r; \quad
h_m \equiv \rho_m + p_m = \rho_m
\]
so that
\begin{equation}
\dot{\rho} +  3 H  h = -
\dot{\Lambda}\,,
\label{conserv2}
\end{equation}
where $\rho = \rho_r + \rho_m$.
The equations (\ref{conserv}) and (\ref{conserv2})
show how energy is transferred from the vacuum to the radiation and
matter densities.  This energy transfer can therefore be understood as
creation of
quanta of radiation and matter from the vacuum. Indeed, employing
the extended form of the first law of thermodynamics suggested in
\cite{edgard1}:
\begin{equation}
d(\rho V) + p d V -  \frac{h}{n} d(n V) = 0 \,,
\label{firstlaw}
\end{equation}
one can connect the evolution of $\rho$ and $p$ with the
evolution of the total number of particles (both photons and massive
particles)  $N = n V$ , where $V$ is
a comoving volume of the observable universe. Since (\ref{firstlaw}) is
equivalent to
\begin{equation}
\dot{\rho} + 3 H h =  h \frac{\dot{N}}{N} ,
\label{firstlaw1}
\end{equation}
comparing with (\ref{conserv2}) gives
\begin{equation}
\frac{\dot{N}}{N} = -  \frac{\dot{\Lambda}}{h} \,.
\label{evolN}
\end{equation}
Therefore in order to create matter or radiation we need
$\dot{\Lambda} < 0$, i.e. $\Lambda$ to decrease with time.

It is clear from this formula that if $\Lambda = const$
we have no particle production and the total number of particles
is conserved. Since radiation and matter do not interact it is
legitimate to assume in this case that radiation and matter evolve
separately according to standard conservation law
\begin{equation}
\dot{\rho_r} +  4 H \rho_r = 0, \qquad
\dot{\rho_m}+ 3 H \rho_m = 0,
\label{conserv3}
\end{equation}
which in conjunction with
\begin{equation}
d(\rho_r V) + p_r d V -  \frac{h_r}{n_r} d(n_rV) = 0 \,, \quad
d(\rho_m V) + p_m d V -  \frac{h_m}{n_m} d(n_mV) = 0 \,,
\label{firstlaw2}
\end{equation}
tell us that the number of photons $N_r = n_r V$ and the number
of  massive particles $N_m = n_m V$ are conserved separately, so
that $N = N_r + N_m = constant$.
\abz

This result gives us a clear understanding that in the case
of $\dot{\Lambda}=0$ the initial vacuum for photons and massive
particles (where $N_r=N_m=0$) will be stable leading to no particle
creation in
the universe. By switching on the source $\dot{\Lambda}\not=0$ in the
right-hand side of the equations (\ref{conserv}) and
(\ref{conserv2}) we effectively switch on
coupling of radiation and matter with gravitational field.
In other words the evolving $\Lambda(t)$  acts as an  interaction of
radiation and  matter with the gravitational field leading,
according to (\ref{evolN}) to creation of photons and massive
particles from vacuum.
\abz
{\it Field equations}
\abz
The field equations
\[
R_{\mu\nu}-{1\over2}Rg_{\mu\nu}
=T^R_{\mu\nu} + T^M_{\mu\nu} + T^Q_{\mu\nu}
\]
are
\begin{eqnarray}
3 H^2 &=& \rho_r  + \rho_m + \Lambda \,, \label{einst0}\\
2 \dot{H} + 3 H^2 &=& - {1\over3}\rho_r  +  \Lambda \,,
\label{einst1}
\end{eqnarray}
and if both are satisfied then the energy conservation equation
(\ref{conserv}) follows identically.

Following \cite{roydavid}, we use $a$ as a dynamic
variable instead of $t$, and consider  the Hubble rate as
$H=H(a)$ (in this case we cannot consider $a=$ constant as a
limiting case for a flat universe). Given $H(a)$, we have two
equations  (\ref{conserv}) and (\ref{einst0}) from which we can
determine $\rho_r(a), \rho_m(a), \Lambda(a)$.

Combining (\ref{conserv}) and (\ref{einst0}) one finds
\begin{equation}
h(a) = - \, a \, [H^2(a)]',
\label{enthalpy}
\end{equation}
where primes denote $d/da$. If we differentiate (\ref{einst1})with
respect
to time and equate with (\ref{conserv}), upon substitution for
$\rho_m$ from (\ref{einst1}) one finds an equation for $\rho_r$.
From this and (\ref{einst1}) we also find an equation for $\rho_m$
\begin{equation}
\rho_r(a) = 3 H^2(a) - R(a) + 3 \Lambda(a) , \qquad \rho_m(a) =
R(a)- 4 \Lambda(a) ,
\label{matter0}
\end{equation}
where
\be
R(a) = 3 a [H^2(a)]' + 12 H^2(a)
\label{ricci}
\ee
is a scalar curvature for metric (\ref{metric}). If we consider
$a=0$ as an initial point of the evolution of the universe, it is
reasonable to assume that the beginning of the evolution
is a vacuum for radiation and matter, i.e.
\[
\rho_r(0)=0\,, \qquad \rho_m(0)=0,
\]
which gives an initial condition
$\Lambda(0) = 3 H_0^2, \, H_0=H(0)$.   Obvious
physical conditions for the energy densities are
\[
\rho_r(a) \geq 0 \,, \qquad \rho_m(a) \geq 0.
\]
Using these conditions, equations (\ref{matter0}) we can obtain the
restriction for $\Lambda$ which holds for
any $a$:
\begin{equation}
a \, [H^2(a)]'(a) \leq \Lambda (a) - 3 H^2(a) \leq
\frac{3}{4} \, a \, [H^2(a)]' \, .
\label{ineq}
\end{equation}

Assuming, that $a \geq 0$, and $H(a) \geq 0$ (i.e. have
expansion of the universe), one can conclude, that this inequality
makes sense only if
\[
H'(a) \leq 0 \,,
\]
i.e. $H$ is a decreasing function of $a$ with the initial value
$H_0$.  Since $\Lambda(0) = 3 H_0^2$, we have that
\[
\Lambda (a) - 3 H^2(a) \leq 0 \,,
\]
i.e. $\Lambda(a) $ is a decreasing function. Therefore we can deduce
from  (\ref{ineq}) the formula for $\Lambda(a)$:
\begin{equation}
\Lambda (a)  = \gamma(a)/2 \, a \, [H^2(a)]' + 3 H^2(a)
\label{lambda}
\end{equation}
where $\gamma(a)$  is a continuous function with the range of
values
\[
\frac{3}{2} \leq \gamma(a) \leq 2 \,.
\]
\abz

On substituting (\ref{lambda}) into (\ref{matter0}) one finds
\abz
\begin{equation}
\rho_r (a) \, = \, \frac{6 - 3 \, \gamma(a)}{2} \; ( - a  [H^2(a)]')\,,
\quad
\rho_m (a)  \, =  \, \frac{4 \, \gamma(a) - 6}{2} \; ( - a [H^2(a)]').
\label{matter1}
\end{equation}
\abz
These  equations satisfy the physical conditions placed on the energy
densities and it follows that if $\gamma = 3/2$,  $\rho_m
\equiv 0$, i.e. we have a universe containing pure radiation;
and {\it vice versa}, and if $\gamma=2$, $\rho_r \equiv 0$, i.e.
we have only matter in the universe.

Now we have the formulas for $\rho_r, \rho_m$ and $\Lambda$ in
terms of $H(a)$ and $\gamma(a)$. The next step hence is to make
some reasonable assumptions about them.
\abz
{\it The form of $\gamma(a)$}
\abz

In order to make our model consistent with the fact that at
present state of the universe radiation and matter evolve
adiabatically, and matter dominates radiation, i.e. according to
standard red shift law $\rho_r \sim a^{-4}$ and $\rho_m \sim
a^{-3}$, we assume that at the late stage of evolution
\[
\frac{\rho_r (a)}{\rho_m (a)}  = \frac{3(2-\gamma(a))}
{2(2\gamma(a)-3)} = \frac{a_{m}}{a},
\]
where $a_{m}$ is the value of the scale factor at time when
the density of radiation and matter are equal ($a_{m} \approx
10^{-1} a_{decoupling}$).  From this formula one can find $\gamma(a)$
for  $a \gg a_m$:
\abz
\begin{equation} \gamma(a) = \frac{6 \left(1
+ a/a_{m}\right)}{4 + 3 a/a_{m}}
\label{gamma}
\end{equation}
\abz

An amazing feature of this formula is that it adequately
describes  a smooth evolution of $\gamma$ from $a=0$ to
$a=\infty$.  Indeed $\gamma(0) = 3/2$, and
$\lim_{a\rightarrow\infty} \gamma(a) = 2$. This result is in a
coherence with the obvious physical expectation that the universe
is dominated by radiation near the beginning of expansion, and
the universe is dominated by matter at the late stage of its
evolution. In other words the formula (\ref{gamma}) gives us a
simple form for $\gamma(a)$ for all values of $a$.
\abz

Assuming now that the simplest form for $\gamma(a)$ for any $a$
is given by (\ref{gamma}), and
plugging (\ref{gamma}) into (\ref{matter1}) and
(\ref{lambda}) we find that
\begin{eqnarray}
\rho_r (a)  &=&  \left[\frac{3}{4 + 3 a/a_{m}} \right]\;
                 [ - a \,  [H^2(a)]']\,, \quad
  \rho_m (a)  = \frac{a}{a_m} \, \rho_r \,, \label{matter2}
\\
\Lambda (a) &=&  \left[ \frac{3 \left(1 +
                  a/a_{m}\right)}{4 + 3 a/a_{m}} \right] \;
                  [a  \, [H^2(a)]']  + 3\, H^2(a).
                  \label{lambda2}
\end{eqnarray}\\
\section{The law of evolution for H(a)}
\abz
{\it The Form  of H(a) in the vicinity a=0}
\abz

To write down the above equations in a form which depends purely
on the scale factor $a$, we need to make predictions about possible
form of $H(a)$. To do this we bring into account two physical
assumptions about the nature of the evolution of the universe in
the vicinity $a=0$ and when $a \rightarrow \infty$. It is by
examining the behaviour of $H(a)$ in the light of these two
assumptions we can predict its possible form.

Let us consider the case when $a \rightarrow 0$. In this case we
can neglect $a/a_{m} \ll 1$ with respect to all  O(1)- quantities
in the expressions (\ref{gamma}), (\ref{matter2}),  and
(\ref{lambda2}) (i.e.   $\gamma \approx
3/2$, corresponding to a purely radiation dominated universe
as $a \rightarrow 0$).

Substituting (\ref{lambda2}) and (\ref{enthalpy}) into
(\ref{evolN}) one obtains the  equation for $N$ when $a
\rightarrow 0$:
\[
\frac{N'}{N} =  \frac{3}{4} \left[ \frac{5}{a} +
\frac{H'}{H} + \frac{H^{''}}{H'}  \right]\,,
\]
which integrates  to
\begin{equation} N =  A \left(- a^5 [H^2]' \right)^{3/4}\,,
\label{numviahubble}
\end{equation}
where $A$ is a constant.
This
expression for $N$ can be rewritten as an equation for $H$:
\begin{equation}
\frac{d}{d a } H^{2}(a) = - {1\over 2A}\, \frac{N^{4/3}(a)}{a^5}\,.
\label{eqhubble}
\end{equation}

We require that initially
there is a standard Minkowski vacuum for radiation, so that
$N(a) \rightarrow  0$ and $n(a) \rightarrow 0$
as $a \rightarrow 0$. This implies
the limiting behaviour
\begin{equation}
N(a) \sim a^{\alpha}\,,~~\alpha > 3\,,~~~~~\mbox{ as }~~
a \rightarrow 0\,.
\label{asymptoteN}
\end{equation}
It then follows from equations (\ref{eqhubble}) and (\ref{asymptoteN})
that
\footnote{It is occurring that the De Sitter stage of the
evolution of the universe appears inevitably  for any
value of $3/2 \leq \gamma \leq 2$ taken as $a \rightarrow 0$,
because, as follows from (\ref{matter2}) and
(\ref{lambda2}), $\Lambda$ dominates evolution. One can show
formally, that taking  $\gamma=$ constant in (\ref{lambda2}), one
can find that $H^2(a) \sim a^{(2\alpha-6)/\gamma} + C$, so that
$H(a) \sim C$ as $a \rightarrow 0$ regardless of the value of
$\gamma$.}
\begin{equation}
H(a) \rightarrow \mbox{ constant}\,,~~~~~\mbox{ as }~~ a
\rightarrow 0\,.
\label{asHzero}
\end{equation}
\abz
{\it H(a) in the vicinity of exit from inflation}
\abz
This form of the asymptotic behavior of $H$ near $a=0$ tells us that
we have a type of expansion {\it a la} inflation. It cannot be an
exact exponential inflation (with  $H=$ constant) for al $a$,
because in this case we would have $\rho_r = \rho_m \equiv 0$,
$\Lambda=$constant, i.e. there would be an eternal stable false
vacuum universe with no production of radiation and matter. Since the
underlying motivation of our model is to obtain the observable
figures for the energy-density and entropy of radiation and
matter in the universe, one must assume that an
initially inflationary universe will evolve into a present-state
universe.
Since inflation is usually understood as an expansion with
acceleration, i.e. with $\ddot{a}>0$, or $H+aH'>0$, and at present
we have decelerated phase, there must be exit
from inflation such that $\ddot{a}=0$,
or $H_e = -a_e H_e', H_e = H(a_e)$. This makes possible to calculate
the $\rho_r, \rho_m, \Lambda$ at $a = a_e$ in terms of $H_e$:
\begin{eqnarray}
\rho_r
(a_e)  &=&  \frac{6}{4 + 3 a_e/a_{m}} \;  H_e^2 \approx
          \frac{3}{2} \; H_e^2 \,, \label{radex} \\
\vspace{4ex}
\rho_m (a_e)  &=&
\frac{6 \, a_e/a_{m}}{4 + 3 a_e/a_{m}} \;  H_e^2 \, \approx
\frac{3}{2} \, \frac{a_e}{a_{m}} \; H_e^2 = \frac{a_e}{a_{m}}
                  \; \rho_r(a_e) \ll \rho_r(a_e)
                  \label{matterex} \\
\vspace{4ex}
\Lambda (a_e) &=&  - \frac{6
\left(1 + a_e/a_{m}\right)}{4 + 3 a_e/a_{m}} \, H_e^2 \,
 + 3\, H^2_e \approx \frac{3}{2} H_e^2.
\label{lambdaex}
\end{eqnarray}

(We assumed  here $a_e \approx 10$ cm, $a_{m} \approx 10^{24}$
cm, and neglected $a_e/a_{m} \approx 10^{-23}$ with respect to
all quantities  of the order $O(1)$).  Comparing (\ref{radex})
and (\ref{matterex}) one can see that the creation of matter
during the inflationary stage is damped with respect to the
creation of photons with the amplitude $10^{-23}$, the universe is
therefore
dominated by radiation at this stage with little matter.

Assuming now that created radiation  is a black-body radiation,
one can estimate the total number of photons created at exit from
inflation.  Using (\ref{radex}) for a standard calculation of
entropy (see \cite{roydavid}) one can show,
that $N_r(a_e) \approx  10^{88}$.
\abz
{\it  $H(a)$ for large $a$}
\abz

In order to find an asymptotic for $H(a)$ for large  $a$ we employ
again the assumption about an adiabatic evolution of radiation and
matter at present, i.e.  $\rho_r  = \alpha a^{-4}$ and
$\rho_m = \beta a^{-3}$. Using the formula (\ref{matter2}) one can
obtain
an equation for $H^2(a)$  for large $a$:
\[
\frac{d}{da} H^{2}(a) =  - \left[\frac{4a_m + 3 a}{3 a a_m} \right] \,
                   \frac{\alpha}{a^4} = - \frac{\alpha}{3 a_m}
                   \left[\frac{4a_m}{a^5} + \frac{3}{a^4} \right]
\]
which integrates to
\[
H^{2}(a) =   \frac{\alpha}{3 a_m}
         \left[\frac{a_m}{a^4} + \frac{1}{a^3} \right]
\]
(We disregard an arbitrary constant of integration because of an
obvious  condition for an open universe that $H \rightarrow 0$ as $a
\rightarrow \infty$).

\abz
{\it  Matching of two asymptotics for $H(a)$}
\abz

We have arrived to two asymptotic formulas for  $H^2(a)$:
\abz
\[
H^{2}(a) \sim \; \; {\rm const} \; \; {\rm as} \quad a \rightarrow 0, \;
\;
{\rm and} \; \;
H^2(a) \sim  \frac{a_m}{a^4} + \frac{1}{a^3} \; \; {\rm as}
\quad a \rightarrow \infty
\]
\abz
Our intention now is to combine these two asymptotics in order
to find a formula for $H(a)$ which is valid for all $a$  and
describes  a smooth transition from inflation to the adiabatic
phase of expansion contingent upon independent evolution of
radiation  and matter.

In accordance with the method of the papers \cite{gmn,infl98},
we try the formula for  $H(a)$ in the following form:
\begin{equation}
H^{2}(a) = C \left[ \frac{a_m}{(b^q + a ^q)^n} + \frac{1}{(d^p + a
^p)^m}
\right]
\label{H2}
\end{equation}
where $q \cdot n = 4$ and $p \cdot m = 3$, as suggested by the
asymptotic behaviour of $H(a)$ examined above.

We can easily fix the constants $b^q$ and  $d^p$ by imposing the
condition for the exit from inflation $2 H_e^2 = - a_e (H_e^2)'$
in (\ref{H2}). We obtain that
\[
b^q = a_e^q, \qquad d^p = a_e^p/2.
\]
(It is interesting to note, that the condition for exit contains
only the products $q n$ and $p m$.)

The final choice of $q, n, p, m$ we will base on the requirement
to have at $a \rightarrow \infty$ the corrections to both leading
terms in (\ref{H2}), i.e. $1/a^4$ and  $1/a^3$ to be $O(1/a^8)$
and $O(1/a^6)$, i.e. we take $q = 4, n = 1, p = 3, m = 1$. The constant
$C$ in (\ref{H2}) can be found from equating both sides at $a=a_e$,
so that finally (compare with a similar result from \cite{caldwell}
and \cite{roydan})
\begin{equation}
H^{2}(a) = H_e^2 \, \frac{6 a_m}{3 a_m + 4 a_e} \,
\left[ \frac{a_e^4}{a_e^4 + a ^4} + \frac{2 a_e^4}{a_m (a_e^3 + 2 a^3)}
\right]
\label{H2f}
\end{equation}
In terms of the dimensionless variable $x=a/a_e$ (\ref{H2f})
can be rewritten in the form:
\begin{equation}
H^{2}(x) = H_e^2 \, \frac{2}{1 + 4/3 \mu} \,
\left[ \frac{1}{1 + x^4} + 2 \mu \frac{1}{1 + 2 x^3}
\right]
\label{H2fx}
\end{equation}
where $\mu = a_e/a_m \approx 10^{-23}$.

This formula describes the overall expansion  of the universe
starting from inflation with  $H(x=0) = \sqrt{2} H_e$. For
$x < 1$ the asymptotic for $H^2$ has a form
\[
H^2 =  \frac{2 H_e^2}{1 + 4/3 \mu} \left[ 1 + 2 \mu - 4 \mu x^3 - x^4 +
\dots
\right]
\]
At exit $H(a_e) = H(x =1) = H_e$. For $x>1$ the asymptotic for
$H^2$ is
\[
H^2 =  \frac{2 H_e^2}{1 + 4/3 \mu} \left[ \frac{\mu}{x^3} +
\frac{1}{x^4}
 - \frac{1}{2}\frac{\mu}{x^6} - \frac{1}{x^8} + \dots
\right].
\]
One can obtain an estimate for $H^2$ at $a=a_m$ (the time when the
matter
and radiation energy densities are equal), or $x = \mu^{-1}$:
\[
H_m \equiv H(x=\mu^{-1}) \approx 2 H_e  \mu^2
\]
\section{Dynamics of radiation and matter}
{\it Calculation  of $\rho_r$, $\rho_m$ and $\Lambda$}
\abz
All further calculations will involve the expression
\[
-a[H^2(a)]' = - x \frac{d[H^2(x)]}{dx} \equiv
 \, \frac{8 H_e^2}{1 + 4/3 \mu} \, \omega(x)
\]
where
\[
\omega(x) =
\left[
\frac{x^4}{(1 + x^4)^2} + 3 \mu \frac{x^3}{(1 + 2 x^3)^2}
\right]
\]
such that
\[
\omega(x=0)=0, \qquad \omega(x=1) = \frac{1}{4} ( 1 + 4/3 \mu).
\]
with the asymptotics for the inflationary period, $x<1$
\[
\omega(x) \approx
 3 \mu x^3 + x^4 - 12 \mu x^6 - 2 x^8 + \dots  \, ,
\]
and  the asymptotic for expansion after inflation, $x>1$
\[
\omega(x) \approx
\frac{3}{4} \frac{\mu}{x^3} +  \frac{1}{x^4} -
\frac{3}{4} \frac{\mu}{x^6} -  \frac{2}{x^8} + \dots  \; .
\]
It is easy to obtain from here
\[
\omega(x = \mu^{-1}) \approx \frac{7}{4} \mu^4.
\]
The expressions for  $\rho_r, \rho_m$  and  $\Lambda$ have the form
\begin{eqnarray}
\rho_r (x)  &=&  \frac{8 H_e^2}{1 + 4/3 \mu} \, \left[\frac{3}{4 + 3 x
\mu} \right]\;
  \omega(x)\,,  \; \; \; \rho_m(x) = x \mu \, \rho_r (x),
                 \label{mattergen} \\
\Lambda (a) &=&  - \frac{8 H_e^2}{1 + 4/3 \mu} \,
                   \frac{3(1+x\mu)}{4+3x\mu}\,
                    \, \omega(x) \,
+ 3 H^2(x).
\label{lambdagen}
\end{eqnarray}
\abz
The leading terms in the asymptotics of these expressions for $x<1$
have the form
\begin{eqnarray}
\rho_r (x)  & \approx&
 \, \frac{6 H_e^2}{1 + 4/3 \mu} \,[ 3 \mu x^3 +  (1-9/4\mu^2) x^4]\,,\;
\;
 \rho_m(x) =  x \mu \, \rho_r (x)\,, \label{radas0}\\
\Lambda (a) &\approx&  \frac{6 H_e^2}{1 + 4/3 \mu} \, [1 + 2 \mu
-7 \mu x^3  - (2+3/4\mu) x^4]
\end{eqnarray}
For the period from the exit from inflation to  the time when
the matter and radiation energy densities are equal i.e. for
$1<x \leq \mu^{-1}$ one finds
\begin{eqnarray}
\rho_r (x)  &\approx&  \frac{24 H_e^2}{1 + 4/3 \mu} \,
\left[\frac{1}{4 + 3 x \mu} \right]\;
 \,  \left[\frac{3}{4} \frac{\mu}{x^3} +  \frac{1}{x^4} \right],
\; \; \; \; \; \rho_m(x) = x \mu \, \rho_r (x)
\label{radasm} \\
\Lambda (x) &\approx& \frac{6 H_e^2}{1 + (4/3) \mu} \, \left\{
- \frac{1+x\mu}{1+(3/4)x\mu}\, \left[ \frac{3}{4} \frac{\mu}{x^3}
+  \frac{1}{x^4}  - \frac{3}{4} \frac{\mu}{x^6} \right] \,
+ \,  \left[ \frac{\mu}{x^3} + \frac{1}{x^4} - \frac{1}{2}
\frac{\mu}{x^6}
\right] \right\}.
\label{lambdaasm}
\end{eqnarray}
On substituting $x = \mu^{-1}$ one gets
\begin{equation}
\rho_r(a=a_m) = \rho_m(a=a_m) = \rho_r (x= \mu^{-1})  \approx
6 H_e^2 \mu^4 =  6 H_e^2 \left[\frac{a_e}{a_m}\right]^4, \;
\label{am}
\end{equation}
\[
\Lambda (a=a_m) = \Lambda(x = \mu^{-1})  \approx 6 H_e^2 \,
 \frac{15}{7} \, \mu^{7} =   \frac{15}{7} \,
\left[ \frac{a_e}{a_m} \right]^7.
\]

From the time when the energy densities were equal till the present
and later  (the matter dominated period) i.e. for $1 \ll \mu x \ll x$
then we find the leading terms are:
\[
\rho_r (x) \approx  6 H_e^2 \frac{1}{x^4}, \qquad
\rho_m (x) \approx  6 H_e^2 \frac{\mu}{x^3}, \qquad
\Lambda (x) \approx \frac{16}{9} \frac{\mu}{x^5}.
\]
These equations describe the standard red shift law $\rho_r \sim a^{-4}$

and $\rho_m \sim a^{-3}$ and show the adequacy of our choice for the
form of $H^2(a)$ above
\abz
\abz
{\it Particle Numbers and specific entropy}
\abz
\abz
Solving, by integration, the equations for $N_r$ and $N_m$ which follow
from
(\ref{firstlaw2})
\begin{equation}
\frac{N_r'}{N_r} = \frac{3}{4} \left[ \frac{\rho_r'}{\rho_r} +
\frac{4}{a}  \right], \qquad
\frac{N_m'}{N_m} =  \left[ \frac{\rho_m'}{\rho_m} +
\frac{3}{a}  \right],
\label{rate}
\end{equation}
one can get  obvious expressions
\begin{equation}
\frac{N_r(a)}{N_r(a_e)} \, = \, \left[\frac{\rho_r(a)}{\rho_r(a_e)
} \right]^{3/4} \, \left[ \frac{a}{a_e} \right]^3, \qquad
\frac{N_m(a)}{N_m(a_e)} \, = \,
\left[\frac{\rho_m(a)}{\rho_m(a_e) } \right] \, \left[
\frac{a}{a_e} \right]^3. \label{numbersviarho}
\end{equation}
One can easily determine  $N_r(a_e)=N_r(x=1)$  from $\rho_r(a_e)$ using
e.g.
the results of \cite{roydavid}. Finally
\be
N_r(x) = N_{r\;e} \, \left[\frac{16 \omega(x)}{4 + 3 \mu x}\right]^{3/4}
\;
         x^3
\label{nr}
\ee
Figure 1 shows the evolution of $y(x) \equiv N_r(x)/N_{r\;e}$ over
the expansion of the universe  The number of quanta of
radiation increases to a maximum value which is given by its
limit as $x \to \infty$
\begin{equation}
N_{r\infty} = N_{re} 2 \sqrt{2},
\label{nradinf}
\end{equation}
(we used $\mu \ll 1$).
\abz

We can not estimate $N_m(a_e)$ from $\rho_m(a_e)$, because we do not
know the mass of the particles constituting matter. Instead of
postulating this mass, we can connect $N_m(a_e)$ with  $N_r(a_e)$
through the asymptotic condition for specific entropy $s$
\begin{equation}
s^{*} \equiv \lim_{x\rightarrow\infty} \frac{N_r(x)}{N_m(x)}
\approx 10^8.
\label{entropy}
\end{equation}
Using  $\rho_m = (a/a_m) \rho_r$, one can
easily prove that
\[
\frac{N_m(x)}{N_{m \, e}} =
\left[\frac{N_r(x)}{N_{re}} \right]^{4/3} =
\frac{16 \omega(x)}{4 + 3 \mu x} \; x^4.
\]
Taking limit $x\rightarrow\infty$ in the last formula one finds
\[
N_{m\infty} = 4 N_{me},
\]
which together with (\ref{nradinf}) and (\ref{entropy}) gives
\[
N_{m e} = \frac{{s^{*}}^{-1}}{\sqrt{2}} \, N_{re}.
\]

From the expressions (\ref{radas0}) and
(\ref{numbersviarho}) one can obtain the asymptotics  for
the number of particles:
\[
N_r(a) \sim a^{21/4},  \qquad N_m(a) \sim a^{7} \qquad {\rm as} \quad
a \rightarrow 0
\]

For the photon creation rate  (\ref{rate})
for $x = a/a_e  < 1$  one finds  from (\ref{radas0}) (using
$\mu=10^{-23}$ and neglecting it with respect to
O(1) terms)  we find
\[
\Gamma_r = \frac{\dot{N_r}}{N_r} = H a \frac{N_r'}{N_r} \approx
\frac{3}{4} H (x)
\left( \frac{21 \mu + 8 x}{3 \mu + x}) \right).
\]
In the limiting case $x \rightarrow 0$ we have
\[
\Gamma_r(x \rightarrow 0) = \frac{21 \sqrt{2}}{4} H_e
\]
Then from  (\ref{rate})  we find that
\begin{equation}
\Gamma_m = \frac{N'_m(a)}{N_m(a)} = \frac{4}{3}\frac{N'_r(a)}{N_r(a)}
= H (x) \left( \frac{21 \mu + 8 x}{3 \mu + x}) \right)
\label{Gammam}
\end{equation}
and
\[
\Gamma_m(x \rightarrow 0) = 7 \sqrt{2} H_e.
\]
\abz
For the specific entropy
\[
s(a) \equiv \frac{N_r(a)}{N_m(a)}
\]
one can easily obtain the equation for its evolution
using (\ref{Gammam}):
\[
\frac{\dot{s}}{s} =  \left[ \frac{\dot{N}_r(a)}{N_r(a)} -
\frac{\dot{N}_m(a)}{N_m(a)}
\right]  = - \frac{1}{3} \, \frac{\dot{N}_r(a)}{N_r(a)} =
- \frac{1}{3} \, \Gamma_r \leq 0.
\]
The decrease of the $s$ is connected with its definition
and with the fact that $N_m$  evolves faster to $0$ than
$N_r$. Since $\Gamma_r \rightarrow 0$ as $a \rightarrow \infty$
we have
\[
\dot{s} \rightarrow 0 \, \; {\rm as} \; \; a \rightarrow \infty
\]
One can find the formula for the evolution of $s(x)$
\[
s(x) = s_e \,
\left[\frac{16 \omega(x)}{4 + 3 \mu x}\right]^{-1/4} \;
         x^3, \qquad s_e \equiv \frac{N_{re}}{N_{me}}
         = s^{*} \sqrt{2}.
\]

\section{Particle creation in decaying cosmology and quantum field
theory
      in curved space}

In this final section we compare our phenomenological
description of matter creation in an expanding universe with some
known results from quantum field theory in curved space. Namely
we compare the results for the $N_r$ presented in a previous section
with the number of particles $N_q$ created from vacuum through quantum
effects  (see  e.g. \cite{birdav,hu,nestott})
\[
\frac{dN_q}{dt} =  c a(t)^3 R^2(t) \qquad {\rm or} \qquad
\frac{dN_q}{da} =  c \frac{a^2 \, R^2(a)}{H(a)} ,
\]
where
$R$ is the Ricci curvature given by the formula
(\ref{ricci}), and the constant "c" takes different values for
different fields.  This formula describes creation of massless
non-conformal particles. We argued in the paper
\cite{nestott} that this formula can be integrated, so that the
total number of
particles created from the initial state at $x=0$ to some $x$
can be presented in the form  ($x=a/a_e$):
\[
N_q(x) = c \int_0^x  \frac{x'^2 \, R^2(x')} {H(x')} dx',
\]
or
\abz
\be
N_q(x) = N_{q \, e} \frac{\int_0^x \frac{x'^2 \, R^2(x')}{H(x')} dx'}
{\int_0^1 \frac{x'^2 \, R^2(x')}{H(x')} dx'}.
\label{nq}
\ee
\abz
where $N_{q \, e}$ is the number of particles created to $a=a_e$
($x=1$).
If we assume now that our species  of particles can be modelled by
quanta, let say,  of  a minimally coupled scalar field, one can
apply  (\ref{nq}) for the universe which evolution is described
effectively   by the formula (\ref{H2fx}). In this case we tacitly
assume that our phenomenological description of transfer of energy
from decaying  $\Lambda$  to radiation and matter can be alternatively
understood as quantum particle creation from vacuum in the metric
(\ref{metric}) with the expansion law (\ref{H2fx}). This assumption
makes sense since, as we have seen above, the most of particles
is created during the inflationary stage, where the contribution
from massive particles can be neglected.

Comparing  two graphs for
$y(x) \equiv N_r(x) / N_{r\,e}$ (formula (\ref{nr}) and
$z(x) \equiv N_q(x) / N_{q\,e}$ (fomula (\ref{nq})
presented at the Fig. 1,
we clearly see that  the behavior of $z(x)$ and
$y(x)$ is quite similar and  the asymptotic values of $z$ and $y$
are of the same  order:
\[
z(\infty) = \frac{N_{q\infty}}{N_{q\,e}} = 1.113..
\qquad \quad y(\infty) = \frac{N_{r\infty}}{N_{re}} =  2 \sqrt{2},
\]
Since $N_{q\,e}$ is a free parameter in th
e formula (\ref{nq})
one can choose it from a natural condition that
\[
N_{q \infty} = N_{r \infty}
\]
so that
\[
N_{q \, e} = 2.533..  \, N_{r\,e}.
\]

We can argue now that our phenomenological theory of production
of radiation and matter via  smooth evolution of the universe from
inflation  to the radiation-dominated stage and then to the dust
stage, which is based on the form  for the Hubble parameter
(\ref{H2fx}), can be treated as a mimic of quantum particle creation
from vacuum. This gives in
a sense a microscopic justification for our phenomenological model.
\newpage
\abz
\unitlength 1mm
\linethickness{0.4pt}
\begin{picture}(150.00,90.00)
\put(10,-35.00){\framebox(150.00,120.00)}
\put(10,65){\special{em:graph mcad}}
\end{picture}
\[ \]
\vspace{15ex}
\begin{itemize}
\item[ ] {\bf Fig. 1} \  \ Graphs for
\[
z(x) = \frac{N_q(x)}{N_{q \, {\rm e}}}, \qquad  \quad
                      y(x) = \frac{N_r(x)}{N_{r\, e}}
\]
in dimensionless units where $x \equiv a/a_e$. The number of
particles created through quantum effects $N_q$ is given by the
formula (\ref{nq}); $N_{q\, e}$ is its value at the time of exit
from inflation. $N_{r}$ and $N_{r\,e}$
are related to photon creation due to decay of $\Lambda$  and
are given by the formula  (\ref{nr}).
It is clearly seen the that the behavior of $z(x)$ and
$y(x)$ is quite similar and  the asymptotic values of $z$ and $y$
ar of the same  order:
\[
z(\infty) = \frac{N_{q\infty}}{N_{q\,e}} = 1.113..
\qquad \quad y(\infty) = \frac{N_{r\infty}}{N_{re}} =  2 \sqrt{2},
\]
\end{itemize}
\section{Conclusion}

We have considered a simple and thermodynamically consistent scenario
encompassing the decay of the vacuum, the creation of radiation
and matter, and a natural smooth transition from inflationary to
radiation- and then matter-dominated expansion.
In order to treat all matter in the universe as created
from a Minkowski vacuum, we impose the physical condition that
the  number of particles is zero in the initial vacuum
state $a=0$. Together with requiring finite particle production
in the observable universe, this constrains the form of
the Hubble rate (\ref{H2fx}), and  gives  an inflationary
universe with smooth exit to the radiation era and then to the dust
era. Using  (\ref{H2fx}) we  calculated then the energy density
for radiation and matter, their partlicle numbers and specific
entropy for the overall  evolution of the universe.

We argued  that our phenomenological theory of production
of radiation and matter via  smooth evolution of the universe from
inflation  to the radiation-dominated stage and then to the dust
stage, which is based on the form  for the Hubble parameter
(\ref{H2fx}), can be treated as a mimic of quantum particle creation
from vacuum initial vacuum and  give  a microscopic justification
for our model.

\end{document}